\documentclass[aps,prd,preprint,tightenlines,groupedaddress,showpacs]{revtex4}
\usepackage{epsf,epsfig,graphics,graphicx}
\usepackage{verbatim,color,ulem}
\begin{document}
\draft

\title{External magnetic fields and the chiral phase transition in QED at nonzero chemical potential}

\author{W.-C. Syu, D.-S. Lee}
\affiliation{Department of Physics, National Dong Hwa University,
Shoufeng, Hualien 974, Taiwan, R.O.C.}


\author{C. N. Leung}
\affiliation{Department of Physics and Astronomy,
University of Delaware, Newark, DE 19716, U.S.A.\vspace{1.0 cm}}

\date{\today}

\vspace*{1.0 cm}
\begin{abstract}
\bigskip
Inspired by recent discussions of inverse magnetic catalysis in the
literature, we examine the effects of a uniform external magnetic
field on the chiral phase transition in quenched ladder QED at nonzero
chemical potential. In particular, we study the behaviour of the effective
potential as the strength of the magnetic field is varied while the chemical
potential is held constant. For a certain range of the magnetic field, the
effective potential develops a local maximum. Inverse magnetic catalysis is
observed at this maximum, whereas the usual magnetic catalysis is observed
at the true minimum of the effective potential.
\end{abstract}

\maketitle


The effect of external fields on the symmetry properties of the
vacuum has been extensively studied in the past decades.
Under the quenched, ladder approximation of QED, chiral symmetry is
dynamically broken at weak gauge couplings when a uniform magnetic
field is present~\cite{GMS,LNA}. The dynamically generated fermion
mass is obtained that increases with growing magnetic field strength.
The phenomenon is referred to as magnetic catalysis of chiral symmetry
breaking. A subsequent analysis~\cite{LW} shows that an improved
truncation beyond the quenched ladder approximation produces a gauge
independent dynamical fermion mass within the lowest Landau level
approximation, provided that it is obtained from the fermion self-energy
evaluated on shell. \\

Recent lattice studies of hot QCD in an external (electro)magnetic field
found that the critical temperature of the chiral phase transition
decreases with increasing magnetic field~\cite{BBE}, contrary to the
expectation from magnetic catalysis. This is known as inverse magnetic
catalysis. Although there are several proposed ideas for explaining this
unexpected behaviour~\cite{FH}, what causes this anomalous phenomenon
remains an open question. Recent field theoretic studies did not arrive
at a definite conclusion about how the critical chemical potential for
the chiral phase transition vary with the strength of the external magnetic
field~\cite{chempot}. \\

Previous works on quenched ladder QED in an external magnetic field show
that chiral symmetry is restored above a certain critical value of temperature
\cite{LLN1,GS,LLN2} as well as critical chemical potential \cite{LLN2}. Since the critical temperature and chemical potential are measured in units of
$\sqrt{|eH|}$, where $H$ is the magnetic field, it suggests that their values  will increase with increasing magnetic field and the system does not exhibit
behaviour of inverse magnetic catalysis. Prompted by the current interest in
inverse magnetic catalysis, we reexamine these earlier works more carefully
to determine the effects of the magnetic field on the chiral phase
transition. We shall focus on the effects on the critical chemical potential
in this paper. \\

We employ the effective potential approach of Ref.~\cite{LMNS}, with
modifications relevant to studying the chiral dynamics in the case of nonzero
chemical potential. The detailed profile of the effective potential and the
location of its extrema will enable one to construct the phase diagram of the
chiral dynamics and understand the nature of the phase transition. \\



We begin by constructing the effective potential for chiral dynamics in terms
of the expectation value of composite local fields,
$\sigma(x)=\langle 0| \bar{\psi}(x)\psi(x)|0\rangle$ and
$\pi(x)=\langle 0|\bar{\psi}(x) i\gamma_5\psi(x) |0 \rangle$. To do so,
consider the generating functional
\begin{eqnarray}
Z[J_{\sigma}, J_{\pi}] &\equiv& \exp{\left(iW[J_{\sigma}, J_{\pi}]\right)}  \nonumber \\
&=& \int {\cal D} \psi(x) {\cal D} \bar{\psi}(x) {\cal D} A_{\mu}(x) \cdot
\nonumber \\
&&\exp{\left( i\int d^4 x\left[{\cal L}+J_{\sigma}(x)\bar{\psi}(x)
\psi(x)+J_{\pi}(x)\bar{\psi}(x)i\gamma_5\psi(x)\right] \right)}\, ,
\label{gen-fun}
\end{eqnarray}
where  ${\cal L}$ is the Lagrangian density of massless QED in a uniform
external magnetic field pointing in the $z$ direction. The expectation values
of the composite fields can be obtained by taking the usual variation of the
generating functional with respect to the sources:
\begin{equation}
\frac{\delta W}{\delta J_{\sigma}(x)}=\sigma(x),\qquad \frac{\delta
W}{\delta J_{\pi}(x)}=\pi(x)\, .
\label{del-W}
\end{equation}
By inverting the expressions (\ref{del-W}) to write $\sigma$ and $\pi$ as a
function of the sources, the effective action can be obtained through the Lengedre transformation
\begin{equation}
\Gamma[\sigma, \pi] = W[J_{\sigma}, J_{\pi}]-\int d^4 x
\left[J_{\sigma}(x)\sigma(x) +J_{\pi}(x)\pi(x)\right] \, ,
\label{eff-act}
\end{equation}
from which
\begin{equation}
\frac{\delta \Gamma}{\delta \sigma(x)}=-J_{\sigma}(x),\qquad
\frac{\delta \Gamma}{\delta \pi(x)}=-J_{\pi}(x).
\label{del-G}
\end{equation}
For spacetime independent fields, $\sigma_0$ and $\pi_0$ are given
by the corresponding constant sources $j_{\sigma}$ and $j_{\pi}$,
respectively. The effective potential is found to be
\begin{equation}
V[\sigma_0,\pi_0]=-\frac{1}{\Omega} \Gamma [\sigma_0, \pi_0] \, ,
\label{effect_pot}
\end{equation}
where $\Omega$ is the spacetime volume. The presence of
chiral symmetry renders the effective action/potential a function of
$\rho=(\sigma^2+\pi^2)^{1/2}$ only. It is thus sufficient and
convenient to simply consider, e.g., the case $\pi=0$ and $\sigma \neq 0$.
The complete functional form of the effective potential can be found via
substituting $\sigma_0=\rho$. In terms of the spacetime independent
generating functional, denoted by $w[j]=W[j]/\Omega$, where we have simplified
the notation by setting $j_{\sigma}=j$, the effective potential now becomes
\begin{equation}
V[\rho]= j \rho -w[j] \, ,
\label{V_fun}
\end{equation}
where
\begin{equation}
w[j]=\int \rho \,\,  d j \, .
\label{w_fun}
\end{equation}
\\

Applying the above to the case of nonzero chemical potential and using the
quenched ladder approximation that takes into account contributions from the
lowest Landau level only, we find
\begin{eqnarray}
 j &~\simeq~& -m_{\mu} + \frac{\alpha}{2 \pi} |eH| \, m_{\mu} \int_{-\infty}^\infty
 dq_3 \int_0^\infty d\hat{q}_\perp^2 ~
  \frac{{\rm e}^{-\hat{q}_\perp^2}}{Q_1 Q_2} \nonumber \\
 & & ~~~~\cdot \left[f_+ (Q_1,Q_2) +
 \theta (Q_1 - \mu ) f_-(Q_1,Q_2)+ \theta( \mu - Q_1) f_- (Q_1,-Q_2)
 \right) \, ,
\label{gapu}
\end{eqnarray}
where $\hat{q}_\perp^2 \equiv (q_1^2 + q_2^2)/(2|eH|)$,
$~Q_1^2 \equiv q_3^2 + m^2_{\mu}$, $~Q_2^2 \equiv q_3^2 + 2
|eH| \hat{q}_\perp^2$, $\alpha$ is the fine structure constant, and $m_{\mu}$
is the infrared dynamical fermion mass which is a function of the chemical
potential $\mu$ ($\mu \equiv |\mu| > 0$) and the external source $j$~\cite{LLN1,LLN2}.  The functions $f_{\pm}$ are defined as
\begin{equation}
f_{\pm} (Q_1,Q_2)=\frac{1}{Q_1 + Q_2 \pm \mu } \, .
\end{equation}
The chiral condensate, $\rho= \langle \bar{\psi} \psi \rangle_{\mu}$,
is found to be
\begin{equation}
 \rho~\simeq~-~\frac{\vert eH
 \vert}{\pi^2} m_\mu \left[\theta(m_\mu-\mu) \int_0^{\sqrt{\vert
 eH \vert}} \frac{dq_3}{\sqrt{q_3^2+m_\mu^2}} + \theta(\mu-m_\mu)
 \int_{\sqrt{\mu^2-m_\mu^2}}^{\sqrt{\vert eH \vert}} \frac{dq_3}
 {\sqrt{q_3^2+m_\mu^2}} \right]_.
 \label{rhoeq}
\label{ccmu}
\end{equation}
\\

\begin{figure}[htp]
\centering
\begin{tabular}{cc}
\includegraphics[scale=0.75]{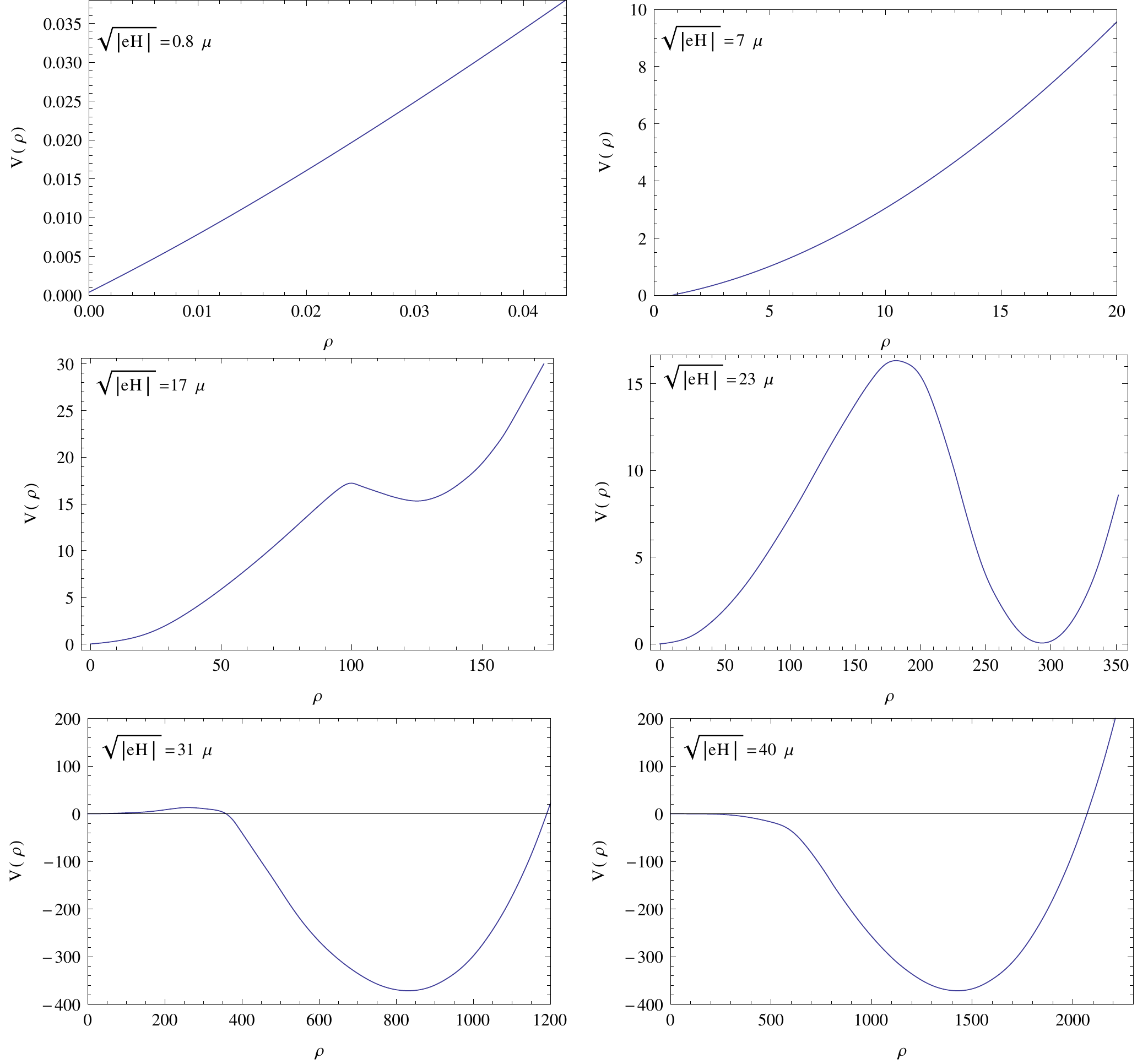}
\end{tabular}
\label{figur}\caption{The effective potential as a function of the
chiral condensate for different values of the magnetic field strength.
Both $V$ and $\rho$ are measured in units of $\mu$, the constant chemical
potential. $\alpha=\pi/10$ for all graphs.}
\end{figure}

The effective potential $V(\rho)$ is evaluated numerically for various values
of the magnetic field in units of the chemical potential $\mu$. Some sample
results are shown in Fig.1. It is found that, for $\sqrt{|eH|} \leq \mu$, the
effective potential has only one minimum, which is located at $\rho=0$,
corresponding to the chirally symmetric phase. See the first graph in Fig.1.
For $\sqrt{|eH|}> \mu$, the second and third graphs in Fig.1 show that, as the magnetic field is increased, the effective potential starts to develop two
additional local extrema, a local minimum and a local maximum, at nonzero
values of $\rho$. The global minimum at $\rho=0$ preserves chiral symmetry. As the magnetic field reaches a certain strength, the two minima become degenerate and a first-order phase transition is about to occur (see the fourth graph in
Fig.1).  When the field strength is above this critical value, the global
minimum of $V(\rho)$ is shifted to a nonzero value of $\rho$ and chiral
symmetry is spontaneously broken. In particular, the expectation value $\rho$
at the global minimum increases with the increase in the magnetic field,
consistent with the expectation of magnetic catalysis (see the last
two graphs in Fig.1). On the contrary, as the strength of the magnetic field
increases, the expectation value at the local maximum, the unstable state,
increases and then shifts toward $\rho=0$. This gives rise to the anomalous
inverse magnetic catalysis effect and is shown in Fig.2. In Fig.3, the
corresponding dynamical mass for the ground state and the unstable state is
also shown. Again the unstable state exhibits features of inverse magnetic catalysis. These figures show again that the magnetic field must exceed a
certain critical value for spontaneous chiral symmetry breaking to take place
when the chemical potential is not zero. \\

\begin{figure}[htp]
\centering
\includegraphics[scale=1]{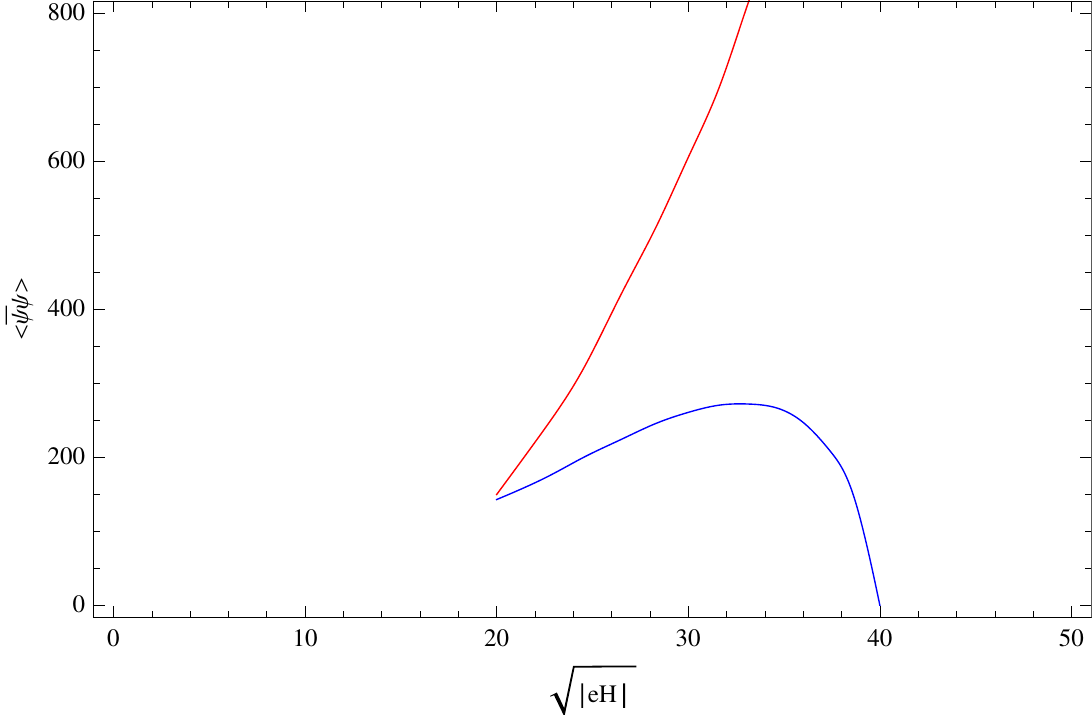}
\label{figur}\caption{The fermion condensate, in units of $\mu^3$, as a
function of the magnetic field ($\sqrt{|eH|}$ in units of $\mu$) for
$\alpha=\pi/10$. The red curve corresponds to the global minimum
of the effective potential and behaves as expected from magnetic catalysis,
whereas the blue curve for the local maximum of the effective potential
exhibits inverse magnetic catalysis behaviour.}
\end{figure}
\begin{figure}[htp]
\centering
\includegraphics[scale=1.3]{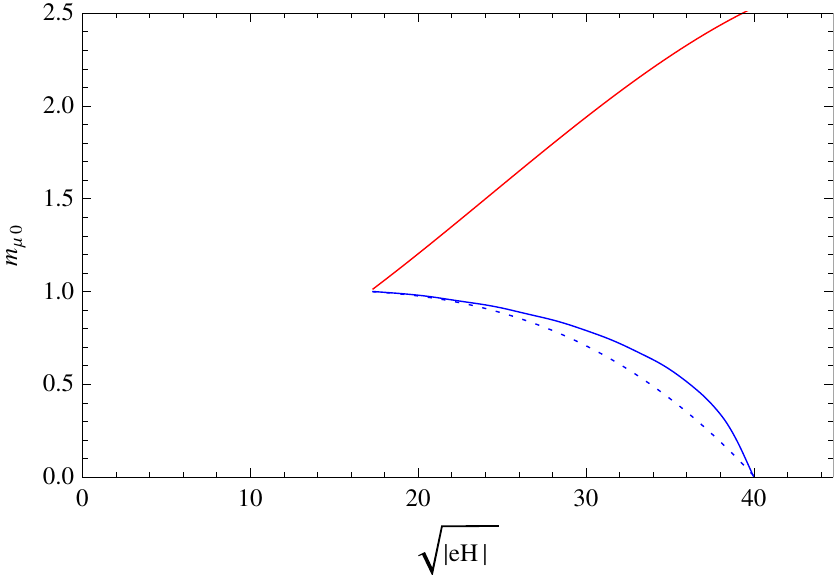}
\label{figur}\caption{The dynamical fermion mass (in units of $\mu$) as a
function of the magnetic field for $\alpha=\pi/10$. The red curve corresponds
to the global minimum of the effective potential and behaves as expected from
magnetic catalysis, whereas the blue curve for the local maximum of the
effective potential exhibits inverse magnetic catalysis behaviour. For
comparison, the approximate result for $m_{\mu 0 ; {\rm unstable}}$ from
Eq.~(\ref{m_unstable}) is shown by the dotted curve.}
\end{figure}


To better understand the numerical results above, let us examine the
behaviour of the effective potential more closely. For $m_{\mu} < \mu$,
the behavior of the effective potential can be found by first
obtaining the corresponding approximate expression of~(\ref{gapu}) for
small $m_{\mu}$,
\begin{equation}
j ~\simeq~ c_1 ~m_{\mu} + c_3~ m_{\mu}^3 + {\cal O}(m_{\mu}^5)\, ,
\label{j_smallm}
\end{equation}
where the constants $c_1$ and $c_3$ depend on the chemical potential
and the magnetic field. Specifically,
\begin{eqnarray}
c_1 (H,\mu)  &=&   -1 + \frac{\alpha}{2 \pi} |eH| \,
\int_{-\infty}^\infty
 dq_3 \int_0^\infty d\hat{q}_\perp^2 ~
 \frac{{\rm e}^{-\hat{q}_\perp^2}}{q_3 Q_2} \cdot \nonumber\\
 && \quad\quad \left[ f_+ (q_3,Q_2) + \theta(q_3 - \mu ) f_- (q_3,Q_2)
 + \theta(\mu - q_3) f_- (q_3,-Q_2) \right] \, ,
\nonumber \\
c_3 (H,\mu) &=& \frac{\alpha}{ \pi} \frac{|eH|}{\mu^2}
 \, \int_0^\infty d\hat{q}_\perp^2 ~
 \frac{{\rm e}^{-\hat{q}_\perp^2}}{2 \vert e H \vert \hat{q}_{\perp}^2
 + \mu^2}\, .
\end{eqnarray}
A nontrivial approximate solution for the gap equation can be obtained
by setting the source $j$ equal to zero, yielding
\begin{equation}
m_{\mu 0; {\rm unstable}} \simeq \sqrt{- c_1 /{c_3}} \, .
\label{m_unstable}
\end{equation}
Note that $c_3$ is always positive and $c_1$ is negative for some range
of the parameters. It can be seen from Fig.3 that this solution, which
is valid for $m_{\mu} < \mu$, corresponds to the unstable local maximum
of the effective potential. We also plotted in Fig.3 this approximate
result for $m_{\mu 0; {\rm unstable}}$ to compare with the exact numerical
result. A reasonably good agreement is found. The decrease of
$m_{\mu 0 ; {\rm unstable}}$ with increasing magnetic field can be
understood by observing that, for a larger value of the magnetic field,
$c_1$ is less negative whereas $c_3$ is larger. \\

Turning now to the behaviour of the chiral condensate $\rho$ for small
$m_\mu$, we find from Eq.~(\ref{rhoeq}) that
\begin{equation}
\rho \simeq -\frac{\vert e H \vert}{\pi^2} m_{\mu} \ln \left[ \left(
\sqrt{\vert e H \vert}+\sqrt{\vert e H \vert +m_{\mu}^2} \right)/
\left(\mu + \sqrt{\mu^2-m^2_{\mu}} \right) \right] \, .
\label{rho_smallm}
\end{equation}
Using the approximate expressions for $j$ and $\rho$ above, one finds
from Eq.~(\ref{w_fun}) that, for $m_\mu < \mu$,
\begin{eqnarray}
w_<&=&\int_0^{m_\mu} \rho (m)\, \frac{dj}{dm} \, dm
\nonumber\\
&\simeq& - \frac{|eH|}{2 \pi^2} \ln \left(\sqrt{|eH|}/{\mu} \right)
 (c_1 m_\mu^2 + \frac{3 c_3}{2} m_\mu^4)
 - \frac{c_1}{16 \pi^2}
 \left[1 + \frac{|eH|}{\mu^2} \right] m_{\mu}^4 + {\cal O}(m^6_{\mu})\, .
\label{w_fun_m=0}
\end{eqnarray}
It follows from Eq.~(\ref{V_fun}) that the corresponding effective potential is
\begin{eqnarray}
V_< &\simeq&  - \, \frac{c_1}{ 2
\pi^2} \vert e H \vert \ln \left(\sqrt{|eH|}/{\mu} \right) m_{\mu}^2
-\left[\frac{3 c_1}{16 \pi^2}
 \left(1 + \frac{|eH|}{\mu^2} \right) \right. \nonumber\\
 && \left. \quad\quad\quad\quad \quad\quad\quad\quad +\frac{c_3}{ 4\pi^2}
 \vert eH \vert \ln \left(\sqrt{|eH|}/{\mu} \right) \right] m_{\mu}^4
 + {\cal O}(m^6_{\mu})\,  .
\label{v_fun_m=0}
\end{eqnarray}
This approximation is compared with the numerical result for the full
effective potential in Fig.4. It is seen to give a reliable description for
$m_{\mu} < 0.5 \mu$. We check that, for sufficiently large magnetic field,
the term proportional to $c_3$ in the coefficient of the $m_\mu^4$ dominates,
thus reproducing the solution (\ref{m_unstable}). The result of (\ref{v_fun_m=0}) also indicates that the unstable local maximum of
the effective potential starts to develop when $\sqrt{|eH|}>\mu$
and $c_1<0$, leading to a positive coefficient for the $m_{\mu}^2$ term and
negative coefficient for the $m_{\mu}^4$ term, and then disappears when
$c_1$ turns positive as the magnetic field exceeds certain critical value,
consistent with our numerical finding in Fig.1. \\
\begin{figure}[h]
\centering
\includegraphics[scale=1.3]{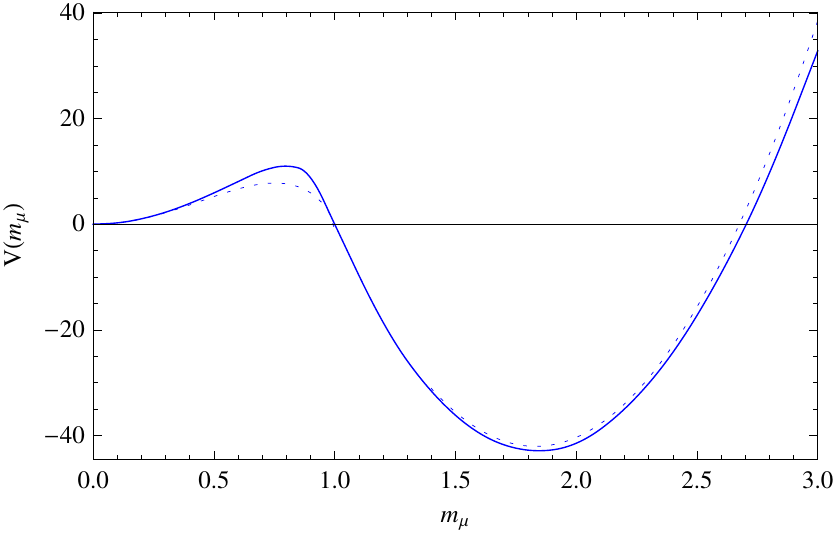}
\label{figur}\caption{Comparison of the full (solid) and the approximate (dotted) effective potentials for $\alpha=\pi/10$ and $\sqrt{|eH|}=28.3~\mu$. Both $V$ and $m_\mu$ are measured in units of $\mu$. }
\end{figure}

We perform a similar analysis for the case $m_\mu > \mu$. Specifically
we examine the behaviour of the effective potential in the neighbourhood
of $m_\mu = m_{\mu 0}$, the ground-state solution of the gap equation.
We find approximate expressions for
\begin{equation}
j \simeq b_1 ~(m_{\mu}-m_{\mu 0})+ b_2 ~(m_{\mu}-m_{\mu 0})^2 +
{\cal O}(m_{\mu}-m_{\mu 0})^3
\label{j_largem}
\end{equation}
and
\begin{equation}
\rho \simeq -\frac{\vert e H \vert}{\pi^2} m_{\mu} \ln
\left[ ( \sqrt{\vert e H\vert}+\sqrt{\vert e H \vert
+m_{\mu}^2})/{m_{\mu}} \right] \, ,
\label{rho_largem}
\end{equation}
from which we obtain the effective potential in the form
\begin{equation}
V_> \simeq v_0 \,+ v_2 \, (m_{\mu}-m_{\mu 0})^2 + {\cal O}
(m_{\mu}-m_{\mu 0})^3 \, .
\label{v_fun_largem}
\end{equation}
The detailed expressions for the coefficients $b_1$, $b_2$, $v_0$,
and $v_2$ are not very illuminating. Instead, we show the result
in Fig.4 and compare it with the full effective potential. The
agreement is quite good and the global minimum of the effective
potential is correctly produced. As shown in Figs.2 and 3, this
true vacuum behaves like what one would expect from magnetic catalysis
as the magnetic field is varied.\\



In summary, we have reexamined the effect of an external magnetic
field on the chiral phase transition in QED at a finite chemical
potential through the effective potential of order parameter fields
relevant to the chiral dynamics. We observe differing behaviour between
the true vacuum (global minimum of the effective potential) and the
unstable local maximum of the effective potential. While both are
solutions to the gap equation, the true vacuum behaves according to
magnetic catalysis while the solution at the local maximum behaves
according to inverse magnetic catalysis. Since the chiral phase
transition is governed by the vacuum solution, we conclude that quenched
ladder QED at nonzero chemical potential exhibits characteristics of
magnetic catalysis. It is useful and interesting to also include
thermal fluctuations in this study. Work along this direction is in
progress. \\

\begin{acknowledgments}
The work of DSL was supported in part by the National Science
Council of Taiwan. Part of the work of CNL was carried out while he
was visiting the Academia Sinica and National Dong Hwa University
in Taiwan. He thanks these institutions for their support and
hospitality.
\end{acknowledgments}

\end{document}